\documentclass[superscriptaddress]{revtex4}
\usepackage{amsmath}
\usepackage{graphicx}
\usepackage{floatflt,epsfig}
\begin{document}

\def\qgp{quark--gluon--plasma}
\def\njl{Nambu--Jona--Lasinio}
\def\cs{chiral symmetry}
\def\ii{\'{\i}}
\def\d{\mbox{d}}

\title{Warm stellar matter with deconfinement: application to 
compact stars}

\author{D.P.Menezes}
\affiliation{Depto de F\'{\i}sica - CFM - Universidade Federal de Santa
Catarina  Florian\'opolis - SC - CP. 476 - CEP 88.040 - 900 - Brazil}

\author{C. Provid\^encia}
\affiliation{Centro de F\ii sica Te\'orica - Dep. de F\ii sica - 
Universidade de Coimbra - P-3004 - 516 Coimbra - Portugal}

\begin{abstract}
We investigate the properties of mixed stars formed by hadronic and quark matter
in $\beta$-equilibrium
described by appropriate equations of state (EOS) in the framework of relativistic mean-field
theory.  We use the non-linear 
Walecka model for the hadron matter and the MIT Bag and the Nambu-Jona-Lasinio 
models for the quark matter. The phase transition to a deconfined quark phase 
is investigated. In particular, we study
the dependence of the onset of a mixed phase and a pure quark phase on the hyperon
couplings, quark model and properties of the hadronic model. 
We calculate the strangeness fraction with baryonic density for the different EOS. With the NJL
model the strangeness content in the mixed phase decreases. 
The calculations were performed for $T=0$ and for
finite temperatures in order to describe neutron and proto-neutron stars. The
star properties are discussed. Both the Bag model and the NJL model predict a mixed phase in
the interior of the star. Maximum allowed masses for proto-neutron stars 
are larger for the NJL model ($\sim 1.9$ M$_{\bigodot}$) than for the Bag
model ($\sim 1.6$ M$_{\bigodot}$).
\end{abstract}

\maketitle

\vspace{0.50cm}
PACS number(s): {21.65.+f, 21.30.-x, 95.30.Tg}
\vspace{0.50cm}

\section{Introduction}

Landau predicted the possible existence of a neutron star after the neutrons were discovered by Chadwick in 1932. In 1934, it was suggested that neutron stars were formed after a supernova explosion, which  happens when the core of a very massive star undergoes gravitational collapse. The first supernova explosion was registered in 1054 by
the Chinese. The Crab Nebula in the Taurus constellation is the 
remnant of this explosion. Recently, a supernova explosion was observed 
in the Magellanic Cloud  170,000 light years from the Earth.
Once the gravitational collapse of a massive star with mass of the order or 
larger than 8 solar masses takes place, a proto-neutron star can be formed.  
Several different stages may happen during the evolution process \cite{prak97,bl86}. The proto-neutron stars are known as evolutionary endpoints and they slowly
cool down to form a neutron star, which is a stable and cold compact star. 
The structure of compact stars is characterized by its mass and radius, which 
in turn are obtained from appropriate equations of state (EOS) at densities about one order of magnitude higher than those observed in ordinary nuclei. At these densities, relativistic effects are certainly important.

In this work we investigate {the equation of state (EOS) of warm, $\beta$-equilibrium
  hadron/quark matter and apply it to determine the properties of mixed stars consisting of hadron matter with
hyperons  and quark matter. In particular we investigate the mixed phase with hadron and
quark matter and search for  the possibility of existence of  a pure quark matter core inside compact stars.
The calculations are performed for finite and zero temperature in order to 
describe proto-neutron  and neutron stars, respectively. We consider only the 
phase after deleptonization, when neutrinos have already diffused out.

For  hadron matter the relativistic non linear Walecka model 
\cite{walecka,bb} with the inclusion of the baryonic octet is used. For sufficiently high
densities it is expected that the formation of  hyperons in a neutron star is energetically
favored. Experimental constraints obtained from hyper-nuclei give estimates of the
hyperon-nucleon and hyperon-hyperon interactions, which impose restrictions on the
expected properties of neutron stars \cite{Glen00}. The appearance of the strange-baryons 
softens the EOS
and lowers the maximum mass of a stable neutron star \cite{Glen00}.  Another  possible source of softening of
the EOS is
the onset of quark matter. According to the hyperon couplings and the properties of the quark
model, this may occur for densities lower or higher than the one corresponding to the onset of
hyperons \cite{prak97}.   
  For 
 quark matter, the MIT Bag model \cite{bag} and the Nambu-Jona-Lasinio model
\cite{NJL} with three quark flavors  were chosen.
The NJL model contains some of the basic
symmetries of QCD, namely  chiral symmetry. It has been very successful in describing the low
lying mesons and predicts at sufficiently high densities/temperatures a phase transition to a
chiral symmetric state \cite{Klevansky92,hk94,PRS87,Ruivo99}. However, it is just an effective
theory that  does not take into account quark confinement. This rises no problems because
we only use the NJL model to describe the deconfined phase.
 At sufficiently large baryonic densities the  quark structure of the
hadrons gives rise to a phase transition into quark matter.
This conversion takes place at densities  a  few times the  nuclear matter density \cite{Glen00}. The Gibb's criteria are
enforced in obtaining the coexistence phase. We also check whether the 
existence of the phase transition depends on the choice of parameters for both 
hadron and quark matter equations of state, as pointed out in \cite{ghosh}. 
The complete EOS is built from the hadronic EOS at small densities, the mixed EOS for intermediate densities and a quark matter EOS for higher densities.

One of the aims of the present work is to understand the importance of 
chiral symmetry for the description of the quark matter. There have been 
contradictory results
presented using the NJL model \cite{scher99,spl00}, namely with respect to the possibility of
existing a mixed phase of hadrons and quarks inside a neutron star. We  
also study the role of the hyperon coupling constants in the appearance of the phase
transition to quark matter and the  change of strangeness composition with density. 
Mixed star properties are obtained by solving the appropriate equations for temperatures
  up to 30 MeV.

The present paper is organized as follows: in section II both models 
used for the quark matter are described and in section III the non linear 
Walecka model with hyperons is reviewed. In both sections the conditions for
beta equilibrium and charge neutrality are discussed. In section IV the mixed 
phase is implemented, the results are shown and discussed and in section V 
the properties of compact stars are computed. Finally, in the last section, 
the conclusions are drawn. 

\section{Quark matter equation of state}

\subsection{The Nambu-Jona-Lasinio Model}
We describe the quark phase within a model with chiral symmetry, the SU(3) NJL model which includes scalar-pseudoscalar 
and the 't Hooft six fermion interaction that  models the axial $U(1)_A$ symmetry breaking. 
The NJL model \cite{Klevansky92,hk94,PRS87,Ruivo99} is defined by the Lagrangian density
\begin{eqnarray}
L\,&=& \bar q\,(\,i\, {\gamma}^{\mu}\,\partial_\mu\,-\,m)\, q +\,g_S\,\,\sum_{a=0}^8\, [\,{(\,\bar q\,\lambda^a\, 
q\,)}^2\,\,+\,\,{(\,\bar q \,i\,\gamma_5\,\lambda^a\, 
q\,)}^2\,]\nonumber\\
&+&\  \,g_D\,\,  \{{\mbox{det}\,[\bar q_i\,(1+\gamma_5)\,q_j]
+ \mbox{det}\,[\bar 
q_i\,(1-\gamma_5)\,q _ j]\, }\},\label{1}
\end{eqnarray}
where $q=(u,d,s)$ are the quark fields and  
$\lambda_a$ $(\,0\,\leq\,a\,\leq\,8\,)$ are the
 U(3) flavor 
matrices. The model  parameters are: 
 $m\,=\, \mbox{diag}\,(m_u\,,m_d\,,m_s\,)$, the  current 
quark mass matrix ($m_d=m_u$), the 
coupling constants $g_S$ and $g_D$ and the cutoff in 
three-momentum space, $\Lambda$.

The set of parameters is chosen in order to fit   
the values in vacuum for 
the pion mass,  the pion decay
 constant,  the
 kaon mass and the quark condensates. We consider the set of parameters \cite{Ruivo99,kun89}: 
 $\Lambda=631.4$ MeV, $ g_S\,\Lambda^2=1.824$,  $g_D\,\Lambda^5=-9.4$, $m_u=m_d=5.6$ MeV and
 $m_s=135.6$ MeV  which where fitted to the following properties: $m_\pi=139$
 MeV, $f_\pi=93.0$ MeV, $m_K=495.7$ MeV, $f_K=98.9$ MeV, $\langle\bar u u\rangle=\langle\bar d
 d\rangle=-(246.7\mbox{ MeV})^3$ and $\langle\bar s s\rangle=-(266.9\mbox{ MeV})^3$.
The frequently used set 
 $\Lambda=602.3$ MeV, $ g_S\,\Lambda^2=1.835$, $g_D\,\Lambda^5=-12.36$, 
$m_u=m_d=5.5$ MeV and 
 $m_s=140.7$ MeV \cite{RKH96} gives similar results but since it has a smaller cut-off and we
 want to study quark matter at high densities we have preferred  the first set.

 The   thermodynamical potential density is given by
${ \Omega}\, = \,{\cal E} -T\,S\,-\,\sum_{i} \mu_i N_i-\Omega_0$,
where the energy density is
 \begin{eqnarray}
{\cal E}\,& =&-2\, N_c \,\sum_i \int 
{d^3 p\over (2\pi)^3}
{p^2 +  m_i M_i\over E_i} \,(n_{i-} - n_{i+})\,\theta (\Lambda^2 -p^2)\,
\nonumber\\
&-&
 2\,g_S\, \sum_{i=u,d,s}  \,\langle\,\bar q_i\,q_i\,\rangle^2
-2\,g_D\,\langle\,\bar u \,u\,\rangle\langle\,\bar d \,d\,\rangle\langle\,\bar s \,s\,\rangle-
{\cal E}_0\,\label{4}
\end{eqnarray}
and the entropy density is
\begin{eqnarray}
S=-2 N_c \sum_{i=u,d,s} \int
{d^3p\over (2\pi)^3}\,\theta (\Lambda^2 -p^2)
\left\{[\,n_{i+} \mbox{ln}(n_{i+}) +
 (1-n_{i+})\, \mbox{ln}(1\,-\,n_{i+})]+ [n_{i+}\rightarrow n_{i-}]\right\}.\label{5}
\end{eqnarray}
In the above expressions $N_c=3$, $T$ is the temperature,  $\mu_i$ ($N_i$) is the chemical potential (number) of
particles of type $i$, and ${\cal E}_0$ and $\Omega_0$ are included in order to ensure ${\cal E}= \Omega=0$ in
the vacuum. This requirement fixes the density independent part
of the EOS. The ground-state of the system
is described by the density 
matrix \cite{Ruivo99} given by
 $f=\,\mbox{diag}\,(f_u\,,\,f_d\,,\,f_s)$ with
\begin{equation}
f_i = {1\over 2}\,[\,I\,(n_{i-}+n_{i+})\,-\,{  \gamma^0\,
M_i\,
+\,{\boldsymbol { \alpha}}\,\cdot\,{\bf p} \over E_i\,}
(n_{i-} - n_{i+})\,]\, \theta (\Lambda^2 -p^2),\label{2}
\end{equation}
where { $I$ is the identity matrix},  $n^{(\mp)}_i$  are the Fermi distribution functions of
  the negative (positive) energy states, 
$n_i^{(\mp)} = [1+ \exp (\mp (\beta\,(E_i \pm \mu_i)))]^{-1}, \quad i=u,d,s.$
In the last equation  $\beta =1/T$,
$M_i$ is 
the constituent quark mass,  $E_i\,=\,(p^2\,+\,M^2_i)^{1/2}$.

The quark condensates and the quark densities are defined, for each one of the 
flavors $i=u,d,s$,  respectively as:  
\begin{equation}
\langle\bar q_i\, q_i\rangle = -2 N_c\, \int {d^3 p\over (2\pi)^3}
{M_i\over E_i}  \,(\,n_{i-}\,-\,n_{i+}\,)\,\theta (\Lambda^2 -p^2), \label{6}
\end{equation}
\begin{equation}
\rho_i\,=\, \langle{q_i}^{\dagger}\, q_i\rangle = 2 N_c\,  \int {d^3p\over (2\pi)^3}\, 
(n_{i-} \,+\, 
n_{i+}\,-1)\,\theta (\Lambda^2 -p^2).\label{7}
\end{equation}

Minimizing the thermodynamical potential $\Omega$ with 
respect to  the constituent quark masses $M_i$ leads to three gap equations for the masses $M_i$
\begin{equation}
M_i\,=\,m_i\,-4\,g_S\,\langle\bar q_i\, q_i\rangle\,-\,2\,g_D\,\langle\bar q_j\, q_j\rangle\langle\bar q_k\, 
q_k\rangle\,,\label{8}
\end{equation}
and cyclic permutations of  $i,\, j,\, k$.

We  introduce an effective dynamical  Bag pressure \cite{bub99,scher99}, 
\begin{eqnarray*}
B&=& 2\, N_c \,\sum_{i=u,d,s} \int 
{d^3 p\over (2\pi)^3}
\left(\sqrt{p^2 + M_i}-\sqrt{p^2 + m_{i}}\right)  \,\theta (\Lambda^2 -p^2)\,\\
&-&
 2\,g_S\, \sum_{i=u,d,s}  \,\langle\,\bar q_i\,q_i\,\rangle^2 
-4\,g_D\,\langle\,\bar u \,u\,\rangle\langle\,\bar d \,d\,\rangle\langle\,\bar s
\,s\,\rangle. 
\end{eqnarray*}
In terms of this quantity the energy density (\ref{4}) takes the form
\begin{equation}
{\cal E}=2\, N_c \,\sum_{i=u,d,s} \int 
{d^3 p\over (2\pi)^3}\sqrt{p^2 + M_i} \,\, (n_{i+}-n_{i-}+1)\, \,\theta (\Lambda^2
-p^2)\,+B_{eff},\qquad B_{eff}=B_0-B,\label{ebageff}
\end{equation}
where $B_0=B_{\rho_u=\rho_d=\rho_s=0}$.  Writing the energy density in terms of $B_{eff}$
allows us to identify this contribution as a Bag pressure and establish a relation with the MIT
Bag model \cite{NJL,bub99} discussed in the next section.

We point out that the NJL model is only valid to describe the quark phase as 
far as the momenta of the quarks are smaller than the momentum cut-off $\Lambda$.

\subsection{The MIT Bag Model}

Quark matter has been extensively described by the MIT Bag model \cite{bag}. 
In its simplest form, the quarks are considered to be free inside a Bag and 
the thermodynamic properties are derived from the Fermi gas model. The
energy density, the pressure and the quark $q$ density are respectively given by:
\begin{equation}
{\cal E}= 3 \times 2  \sum_{q=u,d,s} \int \frac{\d^3p}{(2\pi)^3}
\sqrt{{\mathbf p}^2+ m_q^2} \left(f_{q+}+f_{q-}\right) + Bag,
\end{equation}
\begin{equation}
P =\frac{2}{\pi^2} \sum_{q}
\int \d p \frac{{\mathbf p}^4}{\sqrt{{\mathbf p}^2+m_q^2}} 
\left(f_{q+} + f_{q-}\right) - Bag,
\end{equation}
\begin{equation}
\rho_q= 3 \times 2 \int\frac{\d^3p}{(2\pi)^3}(f_{q+}-f_{q-}), 
\label{rhoq}
\end{equation}
where  $3$ stands for the number of colors, $2$ for the spin degeneracy,
$m_q$ for the quark masses, $Bag$ represents the bag
pressure and the distribution functions for the quarks and anti-quarks are the Fermi distributions
 \begin{equation}
f_{q\pm}=1/({1+\exp[(\epsilon\mp\mu_q)/T]})\;,
\label{distf}
\end{equation}
with $\mu_q$ being the chemical potential for quarks and anti-quarks of type $q$ and
$\epsilon=\sqrt{{\mathbf p}^2+m_q^2}$.
These equations were obtained for finite temperatures. For $T=0$, the 
expressions can be read off the above ones by eliminating the antiparticles and
substituting the particle distribution functions by the usual step functions.

We have used $m_u=m_d=5.5$ MeV, $m_s=150.0$ MeV and $Bag=(180~{\rm MeV})^4$ or
$Bag=(190~{\rm MeV})^4$. { If $m_u,~m_d$ and $m_s$ are chosen as in the NJL model, the behavior of the properties of interest are not altered, since they are more dependent on the Bag pressure than on small differences in the quark
masses.}

\subsection{Quark matter in beta equilibrium}

In a star with quark matter we must impose both beta equilibrium and
charge neutrality \cite{Glen00}.  We consider the stage after 
deleptonization when entropy is maximum and neutrinos diffuse out. In this case the neutrino 
chemical potential is zero.  For $\beta$-equilibrium matter we must add the 
contribution of the leptons as free Fermi gases (electrons and muons) to the 
energy and pressure. The relations between the chemical potentials
of the different particles are given by
\begin{equation}
\mu_s=\mu_d=\mu_u+\mu_e, \qquad  \mu_e=\mu_\mu.
\label{qch}
\end{equation}
For charge neutrality we must impose
$$\rho_e+\rho_\mu=\frac{1}{3}(2\rho_u-\rho_d-\rho_s).$$
For the electron and muon densities  we have
\begin{equation}
\rho_l=2 \int\frac{\d^3p}{(2\pi)^3}(f_{l+}-f_{l-}), \qquad  
l=e,\mu 
\label{rhol}
\end{equation}
where  the distribution functions for the leptons are given in 
eq.(\ref{distf}) by substituting  $q$ by $l$, 
with $\mu_l$  as the chemical potential for leptons of type $l$.
At $T=0$, equation (\ref{rhol}) becomes simply
$\rho_l={k_{Fl}^3}/{3\pi^2}.$

In figure \ref{bag} we plot the effective bag pressure $B_{eff}$ for several temperatures and
for quark matter in beta equilibrium.  As discussed in \cite{scher99} for $T=0$ MeV around
161-163 MeV there is a plateau between $\rho=3 \rho_0$ and $5 \rho_0$. This corresponds to 
 partial chiral symmetry restoration for  quarks $u$ and $d$ but the
chemical potential for quarks $s$, $\mu_s$, is still lower than $m_s$. For higher densities,
$\mu_s>m_s$ and the mass of the quark $s$ goes slowly to its current quark mass value as density
increases.  For high densities, if the current quark masses were zero,  the bag pressure $B\to
0$ and the quarks would behave as a gas of
massless non-interacting  particles inside a large MIT Bag with a bag constant $B_0$
\cite{bub99}. At finite temperature, the value 161 MeV is only reached at  higher densities  and
the plateau slowly disappears.  This behavior is due to the fact that the mass of the quarks
decrease slowlier with $\rho$ and the  quark $s$ exists for $\mu_s<m_s$. In fig
\ref{uds} we show for different temperatures the fractions of $u,\,d$ and $s$
quarks, $Y_i=\rho_i/(N_c\, \rho)$,
 as a function
of $\rho$, the total baryon density. It is clear the effect of temperature on the appearance of
the quark $s$. We will see in section IV that the behavior of the bag pressure with density and
temperature determines the onset of  the mixed phase corresponding to the
coexistence of hadron and quark matter.

\section{Hadronic matter equation of state}

An extension of the NLWM \cite{walecka} is the inclusion of the whole 
baryonic octet ($n$, $p$, $\Lambda$, $\Sigma^+$, $\Sigma^0$, $\Sigma^-$, 
$\Xi^-$,  $\Xi^0$) in the place of the nucleonic sector.
The presence of baryons heavier than the nucleons is expected in 
matter found in the core of neutron stars. The spin 
$\frac{3}{2} \Delta$ and the $\Omega^-$ hyperon should appear at even higher 
densities and hence are not included. The inclusion of other meson fields
describing the $f_0(975)$ and $\phi(1020)$, which are a scalar and a vector 
meson field respectively is also important in reproducing the 
$ \Lambda \Lambda$ interaction \cite{sm} but in this work we restrict 
ourselves to the most common $\sigma$, $\omega$ and $\rho$ mesons.

The lagrangian density of the baryonic octet model (BOM) reads:
\begin{equation}
{\cal L}_{BOM}={\cal L}_{B}+{\cal L}_{mesons}+{\cal L}_{leptons},
\label{octetlag}
\end{equation}
where
$$
{\cal L}_B= \sum_B \bar \psi_B \left[\gamma_\mu\left (i\partial^{\mu}
-g_{vB} V^{\mu}- g_{\rho B} {\mathbf t} \cdot {\mathbf b}^\mu \right) 
-(M_B-g_{s B} \phi)\right]\psi_B,
$$
with $\sum_B$ extending over the eight baryons,
$$g_{s B}=x_{s B}~ g_s,~~g_{v B}=x_{v B}~ g_v,~~g_{\rho B}=x_{\rho B}~ 
g_{\rho}$$
and $x_{s B}$, $x_{v B}$ and $x_{\rho B}$ are equal to $1$ for the nucleons and
acquire different values in different parametrizations for the other baryons,
$$
{\cal L}_{mesons}=\frac{1}{2}(\partial_{\mu}\phi\partial^{\mu}\phi
-m_s^2 \phi^2) - \frac{1}{3!}\kappa \phi^3 -\frac{1}{4!}\lambda
\phi^4
$$
$$
-\frac{1}{4}\Omega_{\mu\nu}\Omega^{\mu\nu}+\frac{1}{2}
m_v^2 V_{\mu}V^{\mu} + \frac{1}{4!}\xi g_v^4 (V_{\mu}V^{\mu})^2
$$
\begin{equation}
-\frac{1}{4}{\mathbf B}_{\mu\nu}\cdot{\mathbf B}^{\mu\nu}+\frac{1}{2}
m_\rho^2 {\mathbf b}_{\mu}\cdot {\mathbf b}^{\mu},
\end{equation}
where
$\Omega_{\mu\nu}=\partial_{\mu}V_{\nu}-\partial_{\nu}V_{\mu} ,
$
$
{\mathbf B}_{\mu\nu}=\partial_{\mu}{\mathbf b}_{\nu}-\partial_{\nu} 
{\mathbf b}_{\mu}
- g_\rho ({\mathbf b}_\mu \times {\mathbf b}_\nu)
$
and ${\mathbf t}$ is the isospin operator.

In the above lagrangian, neither pions nor kaons
are included because they vanish in the mean field 
approximation which is used in the present work and we do not consider 
the possible contribution of pion and kaon condensates.  The electromagnetic field
contribution also vanishes for homogeneous matter and its effect will not be 
taken into account in the mixed phase.
Finally,
\begin{equation}
{\cal L}_{leptons}=\sum_l \bar \psi_l \left(i \gamma_\mu \partial^{\mu}-
m_l\right)\psi_l.
\end{equation}

In the mean 
field approximation, the meson equations of motion read:
\begin{equation}
\phi_0=- \frac{\kappa}{2 m_s^2} \phi_0^2 
-\frac{\lambda}{6 m_s^2}\phi_0^3 + \sum_B \frac{g_s}{m_s^2} x_{s B}~\rho_{s B},
\label{octphi0}
\end{equation}
\begin{equation}
V_0 = -\frac{\xi g_v^4}{6 m_v^2} V_0^3 +
\sum_B \frac{g_v }{m_v^2} x_{v B}~ \rho_B,
\end{equation}
\begin{equation}
b_0 = \sum_B \frac{g_{\rho}}{m_{\rho}^2} x_{\rho B}~  t_{3B}~ \rho_B,
\label{octb0}
\end{equation}
with
\begin{equation}
\rho_B=2 \int\frac{\d^3p}{(2\pi)^3}(f_{B+}-f_{B-}), 
\label{rhob}
\end{equation}
$$
\rho_{s B}= \frac{1}{\pi^2} \int
p^2 \d p \frac{M_B^*}{\sqrt{p^2+{M_B^*}^2}} (f_{B+}+f_{B-}),
$$
with $M_B^*=M_B - g_{s B}~ \phi$,  
$E^{\ast}({\mathbf p})=\sqrt{{\mathbf p}^2+{M^*}^2}$ and
$
f_{B\pm}=
{1}/\{1+\exp[(E^{\ast}({\mathbf p}) \mp \nu_B)/T]\}\;$,
where the effective chemical potential is 
$\nu_B=\mu_B - g_{v B} V_0 - g_{\rho B}~  t_{3 B}~ b_0.$

The energy density  in the mean field approximation reads:
\begin{eqnarray}
{\cal E}&=& 2 \sum_B \int \frac{\d^3p}{(2\pi)^3}
\sqrt{{\mathbf p}^2+{M^*}^2} \left(f_{B+}+f_{B-}\right)+
\nonumber \\
&&\frac{m_s^2}{2} \phi_0^2 + \frac{\kappa}{6} \phi_0^3
+\frac{\lambda}{24}\phi_0^4
+\frac{m_v^2}{2} V_0^2 + \frac{\xi g_v^4}{8} V_0^4 
+\frac{m_{\rho}^2}{2} b_0^2
\nonumber \\
&&+ 2 \sum_l \int \frac{d^3p}{{(2 \pi)}^3} \sqrt{{\mathbf p}^2+m_l^2}
(f_{l+} + f_{l-}),
\label{ener}
\end{eqnarray}
with the leptons distribution functions given after equation (\ref{rhol}).
The pressure becomes
\begin{eqnarray}
P&=&\frac{1}{3 \pi^2} \sum_{B}
\int \d p \frac{{\mathbf p}^4}{\sqrt{{\mathbf p}^2+{M^*}^2}} 
\left( f_{B+} + f_{B-}\right)
\nonumber \\
&&-\frac{m_s^2}{2} \phi_0^2 -\frac{\kappa\phi_0^3}{6} -
\frac{\lambda\phi_0^4}{24}
+\frac{m_v^2}{2} V_0^2 +\frac{\xi g_v^4 V_0^4}{24}
+\frac{m_{\rho}^2}{2} b_0^2 
\nonumber \\
&& + \frac{1}{3 \pi^2} \sum_l \int \frac{{\mathbf p^4} dp}
{\sqrt{{\mathbf p}^2+m_l^2}} (f_{l+} + f_{l-}).
\label{press}
\end{eqnarray}
The expressions given in  this section were obtained for finite temperature, 
but they can be trivially modified for $T=0$ \cite{em}.

\subsection{Hadronic matter in beta equilibrium}

The condition of chemical equilibrium is also imposed through the two 
independent chemical potentials ($\mu_n$ and $\mu_e$) and it 
implies that:
$$\mu_{\Sigma^0}=\mu_{\Xi^0}=\mu_{\Lambda}=\mu_n,$$
$$\mu_{\Sigma^-}=\mu_{\Xi^-}=\mu_n+\mu_e,$$
$$\mu_{\Sigma^+}=\mu_p=\mu_n-\mu_e.$$
For charge neutrality, we must have
\begin{equation}
\sum_B q_B \rho_B + \sum_l q_l \rho_l=0,
\end{equation}
where $q_B$ and $q_l$ stand respectively for the electric charges of baryons and leptons.

For the hadron phase we have used different parametrizations of the non-linear
 Walecka model. Two of them have been  proposed to describe nuclei ground-state properties, 
NL3 \cite{nl3} and TM1 \cite{tm1}, and a third parametrization GL (in the sequel) was proposed
to describe the equation of state of neutron stars \cite{Glen00}.
As shown in the expressions given above, the baryonic octet was included.
In order to fix the meson-hyperon coupling constants we have used
several choices discussed in the literature \cite{Glen00,ghosh}: a) according  to
\cite{gm91,Glen00} we choose the hyperon coupling constants constrained by the binding
of the $\Lambda$ hyperon in nuclear matter, hypernuclear levels and neutron star masses
($x_\sigma=0.7$ and $x_\omega=x_\rho=0.783$) and assume that the couplings to the $\Sigma$ and
$\Xi$ are equal to those of the $\Lambda$ hyperon; b) we take $x_{s B}=x_{v B}= x_{\rho B}=\sqrt{2/3}$ as in \cite{moszk,gl89,cesar,ghosh}. This choice is 
based on quark counting arguments; c) we consider  $x_{s
  B}=x_{v B}= x_{\rho B}=1$ known as universal coupling 
for comparison \cite{glen85}.

 In the results we display, whenever $x_{s B}=x_{v B}= x_{\rho B}$, the unique coupling constant will be called $x_H$. We have 
checked that the parametrizations NL3 and TM1 become unstable respectively for
$\rho/\rho_0=3.4$ and $6.5$ if $x_H=\sqrt{2/3}$ and $\rho/\rho_0=3.5$ and $6.7$ if 
$x_H=1$ owing to the fact that nucleon mass goes to zero. The problem of the effective masses
turning negative has been discussed in \cite{kpe95}, and is due to the fact that the model
includes baryons with different masses and different coupling constants to the $\sigma$
field. 
This behavior is not exhibited if the hyperons are not included, when the 
effective mass of the nucleons  decreases slowly with the density to zero at infinity.  
 However, with the inclusion of the hyperons, and if the hyperon couplings are not chosen
adequately \cite{kpe95}, 
the $\sigma$ field grows much faster with density. For this reason, we have chosen
one of the parametrizations given in \cite{Glen00} (GL) in order to obtain the 
results for the mixed phase discussed in the next section. The chosen 
parameters are $g_s^2/m_s^2=11.79$ fm$^2$, $g_v^2/m_v^2=7.148$ fm$^2$,
$g_{\rho}^2/m_{\rho}^2=4.41$ fm$^2$, $\kappa/M=0.005896$ and 
$\lambda=-0.0006426$, for which the binding energy is -16.3 MeV at the 
saturation density $\rho_0=0.153$ fm$^{-1}$, the symmetry coefficient is 
32.5 MeV, the compression modulus is 300 MeV and the effective mass is $0.7 M$, higher than in
the other two parametrizations.

In reference \cite{sm}, the authors have  imposed a positive value for the 
effective mass by taking its modulus and ignoring the fact that it becomes 
zero and then negative. This gave rise to discontinuities in their figures for some of the parametrizations used. For the TM1 parameter set, there are  no discontinuities, but in their model they  include
other mesons than the $\sigma, \omega$ and $\rho$,  and the behavior of the effective mass also
depends on the hyperon-meson coupling constants used.
 In \cite{scher99} the authors also claim to have used the TM1 
parametrization in order to study the mixed stars, but since no data are given for this
parameter set we  believe it has only been used for stars with no phase transition to the quark
phase.

\section{Mixed phase}

In the mixed phase charge neutrality is not imposed locally but only globally,
{as mentioned in the introduction} 
\cite{Glen00,scher99}. This means that the quark and hadronic phases are not 
neutral separately, but rather, the system will prefer to rearrange itself so 
that 
$$\chi\, \rho_c^{QP}+ (1-\chi) \rho_c^{HP}+\rho_c^l=0,$$  
where $\rho_c^{iP}$ is the charge density of the phase $i$, $\chi$ is the 
volume fraction occupied by the quark phase , $(1-\chi)$ is the
volume fraction occupied by the hadron phase and $\rho_c^l$ is the electric charge density of leptons. We
consider a uniform background of leptons in the mixed phase since Coulomb interaction has not
been taken into account. According to the Gibb's 
conditions for phase coexistence, the baryon chemical potentials, temperatures 
and pressures have to be identical in both phases, i.e.,
$$\mu_{HP}=\mu_{QP},$$
$$T_{HP}=T_{QP},$$
$$P_{HP}(\mu_{HP},T)=P_{QP}(\mu_{QP},T),$$
reflecting the needs of chemical, thermal and mechanical equilibrium, respectively.
As a consequence, the energy density and total baryon density in the mixed 
phase read:
\begin{equation}
<{\cal E}>=\chi\, {\cal E}^{QP}+ (1-\chi) {\cal E}^{HP}+{\cal E}^{l}
\end{equation}
and
\begin{equation}
<\rho>=\chi\, \rho^{QP}+ (1-\chi) \rho^{HP}.
\end{equation}

At this point the EOS in the mixed phase has to be built.
In this work we use for the hadronic phase  the parametrization 
corresponding to $K=300$ MeV and $M^*/M=0.7$ displayed in the previous section. It predicts  a mixed
phase in a range  of densities which depends on the model used  for the quark phase and on the hyperon
coupling constants.
 The higher the hyperon couplings and the 
compressibility and the lower the effective mass the lower lies the range of 
densities corresponding to the mixed phase. This has already been
discussed in \cite{Glen00,ghosh}.
For the quark phase we consider both the NJL model and the MIT Bag model 
with two different bag pressures, as mentioned in section II. The choice of 
the  bag pressures was
done in such a way that for the lower value, $B^{1/4}=$180 MeV, the onset of a quark-hadron mixed phase occurs
before the the appearance of hyperons, and for $B^{1/4}=$ 190 MeV the hyperons appear before the quarks.
With the present study we have concluded that the existence of a quark phase, and  a mixed
  phase is very sensitive to the choice of the hyperon couplings. For a quark phase
  described by the NJL model the mixed/quark  phase only exists if the $\Lambda-\sigma$  coupling is
  not very weak, namely $x_\sigma \geq 0.65$ with the other hyperon coupling constants constrained by the binding
of the Lambda hyperon in nuclear matter. This can be understood in the following way: there will be a quark component
  only if its effect is to soften the EOS. Large hyperon couplings mean that the hadronic EOS
  becomes harder since at high densities the EOS is dominated by the repulsion described by the
  vector mesons. Therefore a transition to a quark phase may be favorable if it softens the
  EOS. A weaker $x_\sigma$ than a $x_\omega$  also favors the phase transition for the same
  reason, i.e., it gives rise to a harder EOS.

In fig \ref{evpt10} we plot the hadronic EOS obtained for several hyperon couplings and the NJL
EOS as a function of the neutron chemical potential $\mu_n$. 
For the hyperon couplings we have considered besides the couplings already introduced in
the present paper, two other sets, both constrained by the $\Lambda$ hyperon binding to nuclear
matter \cite{Glen00}, with $x_\sigma=0.6$ and 0.8. For the first value of  $x_\sigma$ we do not
get a phase transition to quark matter with the NJL model.
The value   $x_\sigma=0.8$ was chosen in ref. \cite{spl00}, but according to
\cite{Glen00}  could be a bit too high if we consider hypernuclear levels which give an
uncertain upper bound of $x_\sigma=0.72$.  The EOS with $x_H$=1. is very
similar to the one obtained with $x_\sigma=0.7,\, x_\omega=x_\rho=0.783$.
 We have assumed
  that the couplings to the $\Sigma$ and $\Xi$ hyperons are equal to those of the $\Lambda$
  hyperon. 
However, we have also tested that if we had taken for the  $\Sigma$ couplings   the  two different sets
  proposed in \cite{kpe95}, which give satisfactory fits to the $\Sigma^-$ atom 
 ($x_{\sigma\Sigma}=0.77,\,x_{\omega\Sigma}=1. ,\,x_{\rho\Sigma}=0.67$
or  $x_{\sigma\Sigma}=0.54,\,x_{\omega\Sigma}=0.67 ,\,x_{\rho\Sigma}=0.67$)  the densities
corresponding to the onset of  the mixed phase and the  pure quark matter phase would not
alter.

According to \cite{bl86} for a newborn neutron star the 
entropy per baryon across the star is
approximately constant, of the order 1 to 2.   This 
corresponds to temperatures which can go up  to 30 MeV \cite{bl86,prak97}. In fig \ref{entropy} we plot the entropy for
different temperatures obtained with  the NJL model. We conclude that, except, for a small
effect in the mixed phase, in the interior of the star ($\rho>2\rho_0$) the 
entropy does not change with temperature. { Exactly the same trend is observed with the Bag model.} 
Therefore, we use the EOS obtained for different parametrizations at fixed
temperatures to determine the properties of proto-neutron stars, and 
{ we are confident that there will not be big differences for not using  
fixed entropies.

 The rise of the entropy in the mixed phase
 corresponds to a decrease in temperature in the interior of the star, if  a fixed entropy calculation were performed. This effect has also been discussed in
ref. \cite{spl00} and it is due to the opening of new degrees of freedom in the system. Similar results were obtained with the Bag model.

In order to implement both phases in a common code, we have imposed the Gibb's
conditions and rewritten the quark chemical potentials of equations (\ref{qch})
in terms of the electron and neutron chemical potentials, i.e.,
$$\mu_u=(\mu_n-2\mu_e)/3 \quad \mu_d=\mu_s=(\mu_n+\mu_e)/3.$$

Examples of the particle populations obtained from the construction of the full EOS are shown in figures \ref{fracoesbag} and \ref{fracoesnjl}. 
Notice that if the Bag model is used, the quarks appear at 
relatively low densities, sometimes lower than the densities for which some of 
the hyperons turn up, depending on the value of $Bag$. This behavior was also observed in 
\cite{Glen00,prak97}. 
If the NJL model is used the appearance of the quarks depends very much on the hyperon coupling constants.  Two different behaviors were obtained: a) the hyperons appear before the quarks
and the mixed phase occurs at much higher densities, ($\sim 5\rho_0$); b) the quarks appear at $T=0$  at lower
densities than  the hyperons if
the effective bag pressure at the onset of the mixed phase is still  well below the plateau
around 162 MeV. In this second case the phase
transition occurs at $\sim 2 \rho_0$ and the effective bag pressure increases 
to its plateau value
for densities inside  the mixed phase. This gives rise to several effects such as a
plateau on  the quark content after the initial increase as can be observed in figure
\ref{fracoesnjl}  at $T=0$. For $T>10$ MeV this last effect disappears because the rise of the
effective bag pressure occurs at higher densities.  With this set of hyperon couplings
there is a transition with temperature: for $T=0$ MeV the quarks appear before the hyperons, at
$T=5$ MeV
both appear at a similar density; for higher temperatures the hyperons appear 
at lower densities.  One main difference between the Bag model and the NJL model
  predictions is the relative fractions of $u,\,d$, and $s$ quarks: in the Bag model  all the
  three types of quarks
  appear at a similar density, then the fraction of $d$ and $s$ quarks increases faster and
  only close to the onset of the pure quark matter the $u$ fraction becomes larger than the
  $s$ quark fraction. At these densities, however, the $u,\,d,\,s$ quark fractions are very
  close to each other, of the order $\sim 0.33$. With the NJL model the $d$ quark appears
  first, at a higher density the $u$ quark appears and still higher the $s$ quark turns up. The
  $d\, (s)$ quark fraction is always the largest (lowest). Only at very high densities, well
  inside the pure quark matter phase, do all three quark fractions get close to 0.33.  These
  behavior is due to the fact that in the NJL model, only at quite  large densities the mass of the $s$ quark gets close to its current mass value.

For all EOS obtained  the higher the temperature, the  lower the densities for
which the hyperons appear. The onset of the mixed phase occurs in most cases studied at 
  higher densities for a higher  temperature, the exception being the 
 the NJL model for $x_H=\sqrt{2/3}$. This can be confirmed in 
 Fig. \ref{fulleosbag}  
where the full EOS, obtained respectively with the
Bag and the NJL models, for $x_H=\sqrt{2/3}$, are given
for different temperatures. For this choice of hyperon couplings 
the Bag model predicts a  mixed phase  at lower
densities (energies). 
 For 
 the other two choices of hyperon couplings studied with the NJL model, 
the mixed
phase occurs before the effective bag pressure reaches the 163 MeV plateau, and this dictates
the difference. For $x_s=0.7$ and $x_H=1.0$ the onset of the mixed phase is determined by the low density behavior of the
  NJL EOS. As discussed in \cite{hanaus01}, the EOS obtained within this parametrization of NJL
  model
   contains mechanical unstable regions which give the possibility of a first-order phase
   transition of the liquid-gas type. Depending on the set of hyperon couplings chosen, it can
   happen that the mixed phase appears precisely at the density above which the NJL is stable
   against density and charge fluctuations.

The presence of strangeness in the core and crust of neutron and proto-neutron
stars can have important consequences in  understanding  some of their properties
\cite{alcock86,blc99,lat98,pons01}. We have calculated the strangeness content of the different EOS as a function
of density. In figures  \ref{cargabag} and \ref{carganjl} we plot the 
strangeness fraction $r_s$ defined as
\begin{equation}
r_s=\chi \, r_s^{QP}+ (1-\chi)\, r_s^{HP}
\end{equation}
with
$$r^{QP}_s=\frac{\rho_s}{3\rho}, \quad r^{HP}_s=\frac{\sum_B |q^B_s|\rho_B}{3\rho}, $$
where $q_s^B$ is the strange charge of baryon $B$.
When the quark phase is described by the Bag model the strangeness fraction rises steadly and
at the onset of the quark phase it has almost reached  1/3 of the baryonic matter.  This behavior is independent of the hyperon-meson coupling constants used in this work.
 The NJL model predicts a
different behavior: in the mixed phase  the strangeness  fraction decreases and increases
again for pure quark matter. This behavior is due to fact that for the densities at which the
mixed phase occurs the mass of the strange quark is still very high. The overall effect of
temperature is to increase the strangeness fraction, except for the mixed phase  with
the NJL model where  the strangeness fraction decreases more   strongly for higher temperatures.

\section{Mixed Star Properties}
In the present section we investigate  the properties of stellar objects formed by matter described by
the EOS studied in the previous sections.
Mixed proto-neutron and neutron star profiles can be obtained from all the EOS 
studied by solving the Tolman-Oppenheimer-Volkoff (TOV) equations \cite{tov}, 
resulting from a simplification of Einstein's general relativity 
equations for spherically symmetric and static stars. 
For a certain EOS the TOV equations for pressure and mass are integrated from 
the origin for a set of arbitrary choices for the the central density, in such
a way that they define a one-parameter family of stars.
The neutron star mass is an increasing function of its central density until 
it reaches a maximum, where matter can collapse to a black hole. It means that 
the values of the maximum mass of a star is very important in the determination
of the mass of possible black holes. The Oppenheimer-Volkoff limit gives the 
maximum value of the neutron star mass. There are theoretical previsions that
the maximum mass varies from 0.72 solar mass ($M_{\odot}$) obtained for pure
neutron matter stars up to 3.26 $M_{\odot}$ \cite{chung}. 
Other authors \cite{cesar}, based on recent calculations, claim that they can be in the interval 1.6 
$M_{\odot}$ and 2.1 $M_{\odot}$. According to \cite{Glen00}, the observed values lie between
1.2 $M_{\odot}$ and 1.8 $M_{\odot}$. 

In tables 1 and 2  we show the values obtained for the maximum mass  of a neutron star as 
function of the central density for some of the EOS studied in this work and  for 
different temperatures. 
In table 1 the GL force and the NJL model were used to derive the full EOS, 
while in table 2 the EOS was obtained from the GL force and the MIT Bag model.
Several conclusions can be drawn. For all EOS studied, the central energy density $\varepsilon_0$ falls inside the
  mixed phase. In some cases, at finite temperature, it can even occur in the quark
  phase. This happens   in the calculation with NJL if the universal coupling is chosen  and in
  the calculation with the Bag model with $B^{1/4}=180$ MeV. Similar results were obtained in
  \cite{spl00}. However, the present results are different from the ones discussed in
  \cite{scher99}, where the parametrization of the EOS used allowed for the appearance of the
  mixed phase only in a quite narrow mass range. 
The maximum masses of the stars do not show any regular behavior with temperature, contrary to
the steady increase of the maximum allowed mass obtained with the GL model alone as shown in table 3. It is worth emphasizing that the EOS for T=0 and T=10 MeV are very close to each other (at least for GL plus Bag models), what could have led to very small differences in the solutions of the TOV equations.
The
trend seems to be a small decrease at lower temperatures followed by a small increase. Only in
one case we have obtained a continuous increase of the mass. 
A discussion of the possibility of a back
hole formation after the deleptonization and cooling of the proto-neutron star requires a more
careful study. In principle, this could happen if the maximum allowed mass at $T=0$ MeV is lower
than the maximum  allowed mass at finite $T$.

Comparing tables 1 and 2 we conclude  that the Bag model allows for smaller maximum masses, of
the order $\sim 1.6\,\, \mbox{ M}_{\bigodot}$, than the
NJL model, $\sim 1.9 \,\, \mbox{ M}_{\bigodot}$. Several quite high mass limits have been recently
determined \cite{hhj,hh00} which could rule out soft EOS with phase transition to deconfined
matter. In particular, the determination of the pulsar Vela X-1 gives the mass M$_X= 1.86\pm 0.16 \,\mbox{ M}_{\bigodot}$.  However, these high mass limits do not exclude a phase transition to quark matter within the
NJL model. It is also seen that the maximum mass does not depend much  on the hyperon couplings
although in the NJL the same is not true for  the onset  of the mixed phase. 
 For comparison we
show in table 3 the properties of stars obtained with the GL model (transition to the quark
phase not included).  While the maximum masses obtained with Bag model are $\sim$ 15 \% smaller
than with the hadron EOS, the ones  obtained within the NJL model are only $\sim$ 3\%
smaller. 

We will not discuss the radius of the maximum mass star because it is sensitive to the
low density EOS and we did not describe properly this range of energy densities.

\section{Conclusions}

In the present paper we have studied the EOS for proto-neutron stars using 
both the Bag model and
the NJL model for  describing the quark phase and a relativistic mean-field description in
which baryons interact via the exchange of $\sigma-,\, \omega-,\,\rho-$ mesons for the hadron phase.
Parametrizations TM1 and NL3 fitted to the ground-state properties of stable and unstable nuclei
proved to be inadequate because, due to the inclusion of hyperons,  the nucleon mass becomes
negative at relatively low densities. We have considered a parametrization which describes the
properties of saturating nuclear matter proposed in \cite{Glen00}. For the hyperon couplings we
have used three choices and verified that for the NJL model the onset of the mixed phase is  sensitive to the hyperon couplings.  
We have verified that  the choice of  other sets of hyperon couplings, namely couplings to the
  $\Sigma$ and $\Xi$ different from the $\Lambda$ hyperon, did not have any effect on the onset
  of the mixed phase.

If the quark phase is described by the Bag model the strangeness content of
the EOS rises steadly and
at the onset of the quark phase it has almost reached  1/3 of the baryonic matter. 
However,  the NJL model predicts a
different behavior: in the mixed phase  the strangeness  fraction decreases and increases
again for pure quark matter.  The overall effect of
temperature is to increase the strangeness fraction, except for the mixed phase  with
the NJL model where  the strangeness fraction decreases more   strongly for higher temperatures.
The strangeness content will influence the cooling rates of the star \cite{lat98,pons01}.

The consequences of using a model with chiral symmetry have been discussed. The onset of
  the mixed phase is the result of a delicate equilibrium between the hyperon couplings and the
  rise of the effective bag pressure due to partial restoration of the chiral symmetry.  The
  value of the  effective bag pressure also dictates whether the onset of hyperons occurs at
  larger or lower densities than the onset of the mixed phase. Temperature will alter the
  order if at $T=0$ MeV the onset of the mixed phase occurs at lower densities.   A large
  $s$ quark mass until very high density gives rise to a decrease of the strangeness content of the 
  EOS in the mixed phase. In the quark phase the fraction of the $s$ quark 
becomes  similar to the
  $d$ and $u$ quark fractions at densities greater then $10 \,\rho_0$. 

The appearance of the mixed phase in $\beta$-equilibrium matter at a fixed temperature
  gives rise to an increase of the entropy per baryon due to the opening of 
other allowed
  channels.  This will  affect neutrino-matter interaction rates \cite{lat98}. 

With the EOS studied in the present work,  the mixed stars obtained from the integration of the TOV
equations  contain a core constituted by a mixed phase. Only in very special cases, a small core of pure quark matter exists. It was shown that the maximum masses of mixed stars obtained
with the Bag model are of the order of $\sim 1.6$ M$_{\bigodot}$,  smaller than  the maximum
masses obtained within the NJL model, $\sim 1.9$M$_{\bigodot}$.  Even with a transition to
a deconfinement phase the masses predicted can be quite high. Maximum masses for neutron stars
obtained in the present work are larger than the ones obtained in \cite{prak97}.
The effect of temperature in the maximum masses allowed is not strong.  The central energy densities,
though, decrease with temperature.

We are aware of the importance of neutrino trapping in the description of 
proto-neutron star properties and this feature will be incorporated in a 
forthcoming work. The effects arising from the inclusion of pion 
and kaon condensates seem not to be very large and also very dependent of the
choice of the hyperon to meson coupling constants \cite{Glen00,prak97}, which 
are not known quantities. For these reasons they were not incorporated in this
 study. Some authors \cite{sm,pmzsg} have claimed that other meson 
fields, namely the scalar meson field $f_0(975)$ and the vector meson field
$\phi(1020)$, should also be incorporated in order to reproduce the observed 
strongly attractive $\Lambda \Lambda$ interaction. We 
believe that 
the inclusion of these meson fields will not have a great influence on the 
properties of the  stars under consideration.
It is well known that non-rotating neutron stars are practically undetectable. The extension of this work in order to include the rotation mechanism
discussed in \cite{hhj,Glen00} is under consideration. 

Moreover, the quark-meson coupling
model (QMC) \cite{guichon}, which describes nuclear matter as a system of 
non-overlapping MIT Bags which interact 
through the effective scalar and vector mean fields, is also a good option for
the description of the hadron phase. The quark phase appears naturally as
the breakdown of the confining bags takes place. This investigation is also 
under way. In the present work we have only  considered  unpaired quark matter.However, recent studies seem to show that the quark phase, if present, would 
be in a color superconductor phase \cite{scs,others}.  The possible existence 
of a color superconductor quark  phase in compact
stars has been studied in  references \cite{ar} with  the Bag model and 
\cite{bub03} using the
NJL model. In general,  it was shown that the inclusion of a
color superconductor phase would soften the EOS at lower pressures 
giving rise to the onset of quarks, either through
a sharp transition or a mixed phase,  at lower densities.  With the 
NJL model and a sharp transition assumption no stable stars with a quark core 
were found. Properties of compact
stars obtained within the Bag model depend on the parameter set used and 
the sharp/mixed transition considered, but it was shown that color 
superconductivity could boost the mass of
the star.  Further studies on the possible existence and consequences of 
a color superconductor phase are still needed.

\section*{ACKNOWLEDGMENTS}
We would like to acknowledge fruitful discussions with C\'elia Sousa and Marcelo Chiapparini.
This work was partially supported by CNPq (Brazil), 
CAPES(Brazil)/GRICES (Portugal) under project 003/100 and FEDER/FCT (Portugal) under the project .
  POCTI/35308/FIS/2000.

\newpage

\begin{center}
Table 1 - Mixed star properties for the EOS obtained with the GL force and the
NJL model.
\end{center}
\begin{tabular}{lcccccc}
\hline
&T (MeV)  & $M_{\mbox{max}}/M_{\bigodot}$& 
$\varepsilon_0$ (fm$^{-4}$) & $\varepsilon_{min}$ (fm$^{-4}$) & $\varepsilon_{max}$ (fm$^{-4}$)\\
\hline
$x_H=\sqrt{2/3}$ & 0& 1.84& 6.29 &4.60&7.25\\
&10&  1.83 &  6.34 & 4.58&7.14\\
&20&  1.84& 6.26 &4.50 &6.66\\
&30&  1.85& 5.85 &4.23 &5.84\\
\hline
$x_s=0.7$ & 0&1.90    & 5.01&1.30&5.21\\
$x_\omega=0.783$&10& 1.89 &4.98  &  1.31&5.13\\
$x_\omega=x_\rho$&20& 1.89  &4.81  &1.37&4.82\\
&30& 1.90  & 4.34 &1.61&4.44\\
\hline
$x_H=1.$&0 & 1.92 & 5.08 & 1.30 & 5.22\\
&10. &  1.89 & 5.13 &  1.54 & 5.10\\
&20 &  1.88 &  4.92 & 2.20 & 4.83\\
&30 &  1.87 &  4.58 & 2.18& 4.46\\
\hline
\end{tabular}

\vspace{0.5cm}
\begin{center}
Table 2 -  Mixed star properties for the EOS obtained with the GL force and 
the MIT Bag model.
\end{center}
\begin{tabular}{lcccccc}
\hline
& T (MeV)  & $M_{\mbox{max}}/M_{\bigodot}$& 
$\varepsilon_0$ (fm$^{-4}$) & $\varepsilon_{min}$ (fm$^{-4}$) & $\varepsilon_{max}$ (fm$^{-4}$)\\
\hline
 Bag$^{1/4}$=180 MeV & 0  & 1.40 &  7.38 & 1.17 & 4.62\\
$x_H=\sqrt{2/3}$ & 10  & 1.39 &  7.05 & 1.17 & 4.58\\
& 20  & 1.42 &  4.18 & 1.20 & 4.47\\
\hline
Bag$^{1/4}$=190 MeV & 0   & 1.64  & 4.58 & 1.81 & 6.06 \\
$x_H=\sqrt{2/3}$ & 10  & 1.59  & 4.86 & 1.85 & 6.03 \\
& 20  & 1.67 &  4.18 & 1.90 & 5.93 \\
& 30  & 1.76  & 3.18 & 1.96 & 5.78 \\
\hline
$x_s=0.7$ & 0 & 1.63    & 4.43 &1.53&6.0\\
$x_\omega=0.783$&10 & 1.63  &4.41  & 1.57 &5.95\\
$x_\omega=x_\rho$&20 & 1.63   & 4.10  &1.62&5.86\\
&30 & 1.64   & 3.82 &1.67&5.69 \\
\hline
$x_H=1$ & 0   & 1.64  & 4.49 & 1.63 & 6.01 \\
&10  & 1.65 & 4.40 & 1.66 & 5.98 \\
&20  & 1.67  & 4.12 & 1.71 & 5.87 \\
&30  & 1.72  & 3.61 & 1.76 & 5.70 \\
\hline
\end{tabular}

\vspace{0.5cm}
\begin{center}
Table 3 -  Compact star properties for the EOS obtained with the GL force
\end{center}
\begin{tabular}{lccccccc}
\hline
&T (MeV) & $M_{\mbox{max}}/M_{\bigodot}$&  $\varepsilon_0$ (fm$^{-4}$) \\
\hline
$x_H=\sqrt{2/3}$ & 0 & 1.93 & 6.36 \\
& 10 & 1.93 &  6.34 \\
& 20 & 1.95 &  6.20 \\
& 30 & 1.97 &  5.85 \\
\hline
\end{tabular}

\begin{figure}
\begin{center}
\includegraphics[width=9.cm]{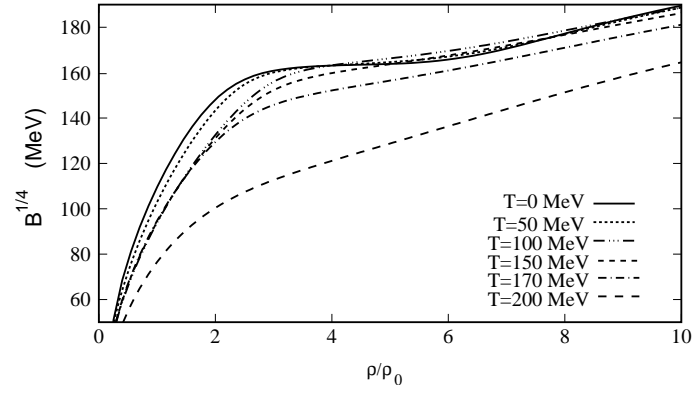}
\end{center}
\caption{NJL model in beta equilibrium:  the bag effective pressure at different temperatures}
\label{bag}
\end{figure}

\begin{figure}
\begin{center}
\includegraphics[width=9.cm]{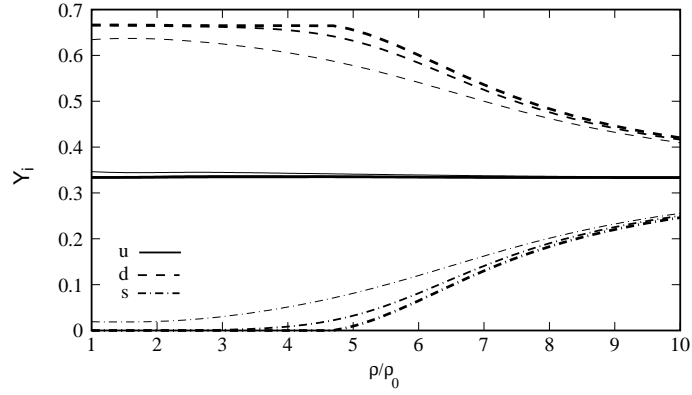}
\end{center}
\caption{The fraction of quarks u,d,s  in beta equilibrium at T=0,  20 and 50 MeV (from very thick to thin)}
\label{uds}
\end{figure}
\begin{figure}
\begin{center}
\includegraphics[width=9.cm]{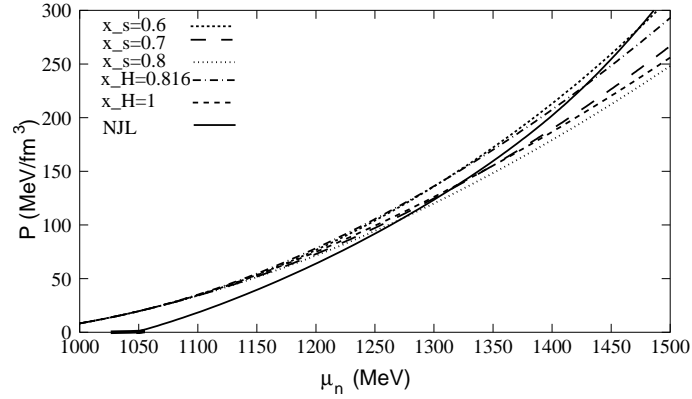}
\end{center}
\caption{The EOS for the NJL model and for GL model with different choices of the hyperon couplings)}
\label{evpt10}
\end{figure}

\begin{figure}
\begin{center}
\includegraphics[width=9.0cm]{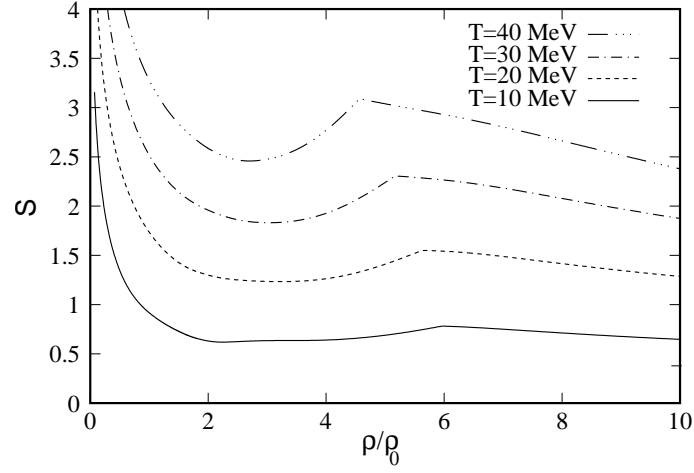}
\end{center}
\caption{Entropy per baryon for different temperatures. The quark phase is described by the NJL model.}
\label{entropy}
\end{figure}

\begin{figure}
\begin{center}
\begin{tabular}{cc}
\includegraphics[width=7.cm]{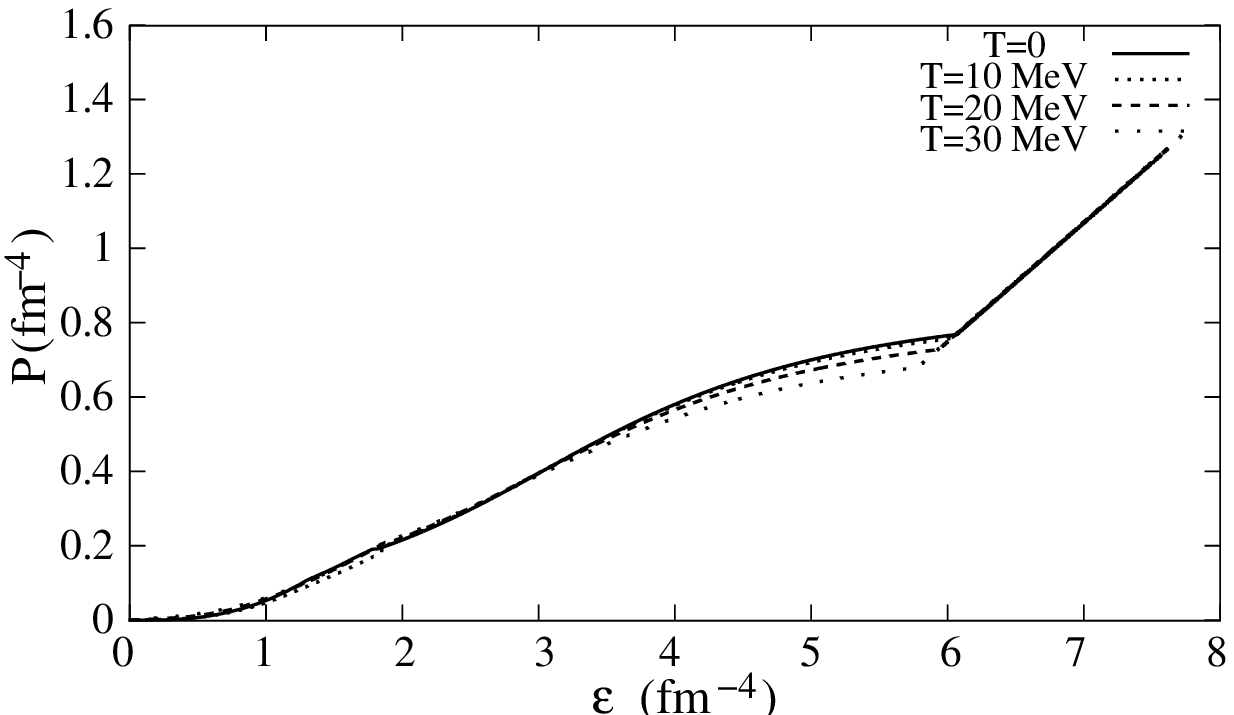}&
\includegraphics[width=7.0cm]{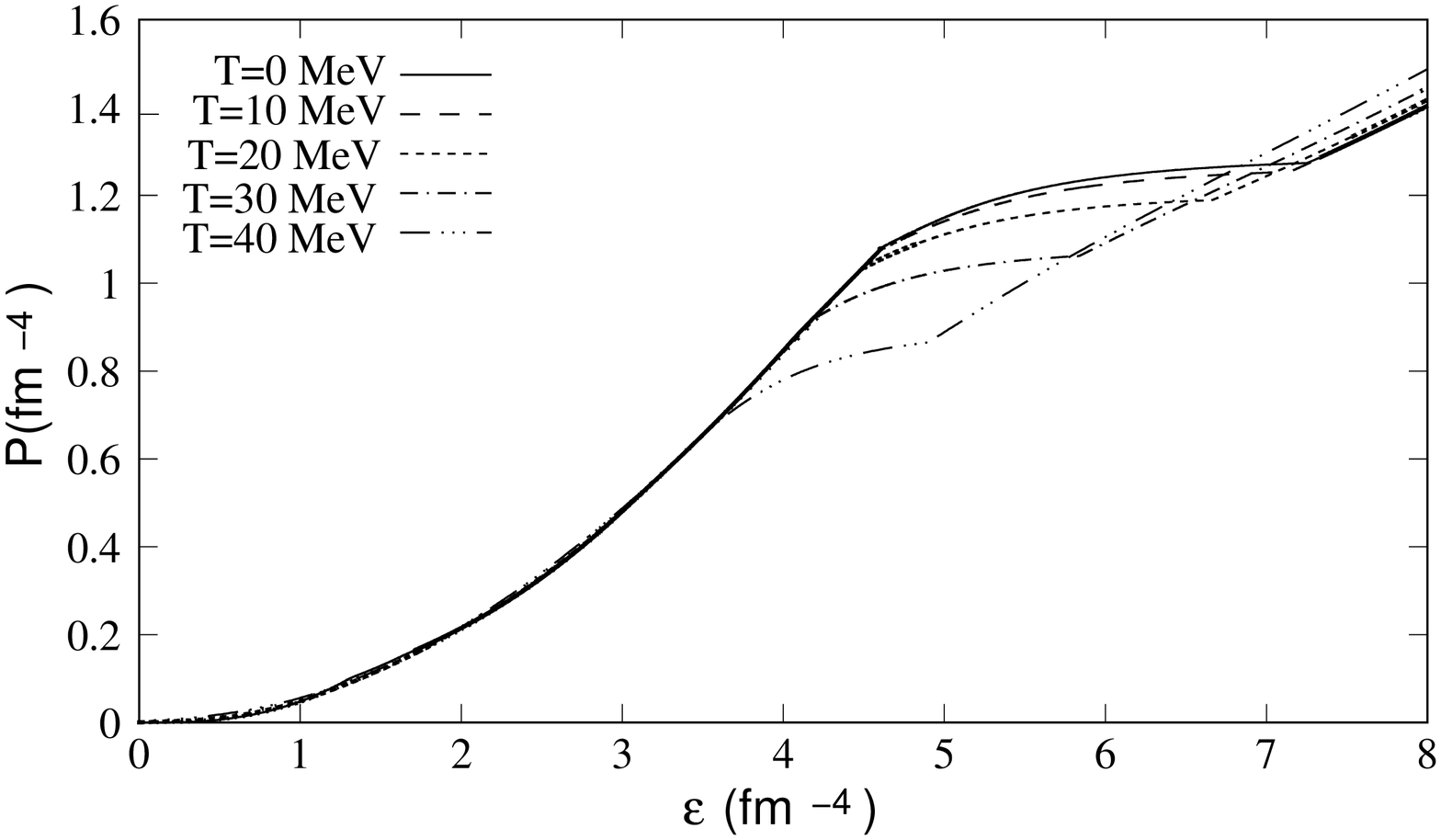}\\
a)&b)
\end{tabular}
\end{center}
\caption{EOS obtained with the GL force plus a) Bag model for Bag$^{1/4}$=190 
MeV; b) NJL model. In both cases $x_H=\sqrt{2/3}$.}
\label{fulleosbag}
\end{figure}

\begin{figure}
\begin{center}
\begin{tabular}{cc}
\includegraphics[width=7.0cm]{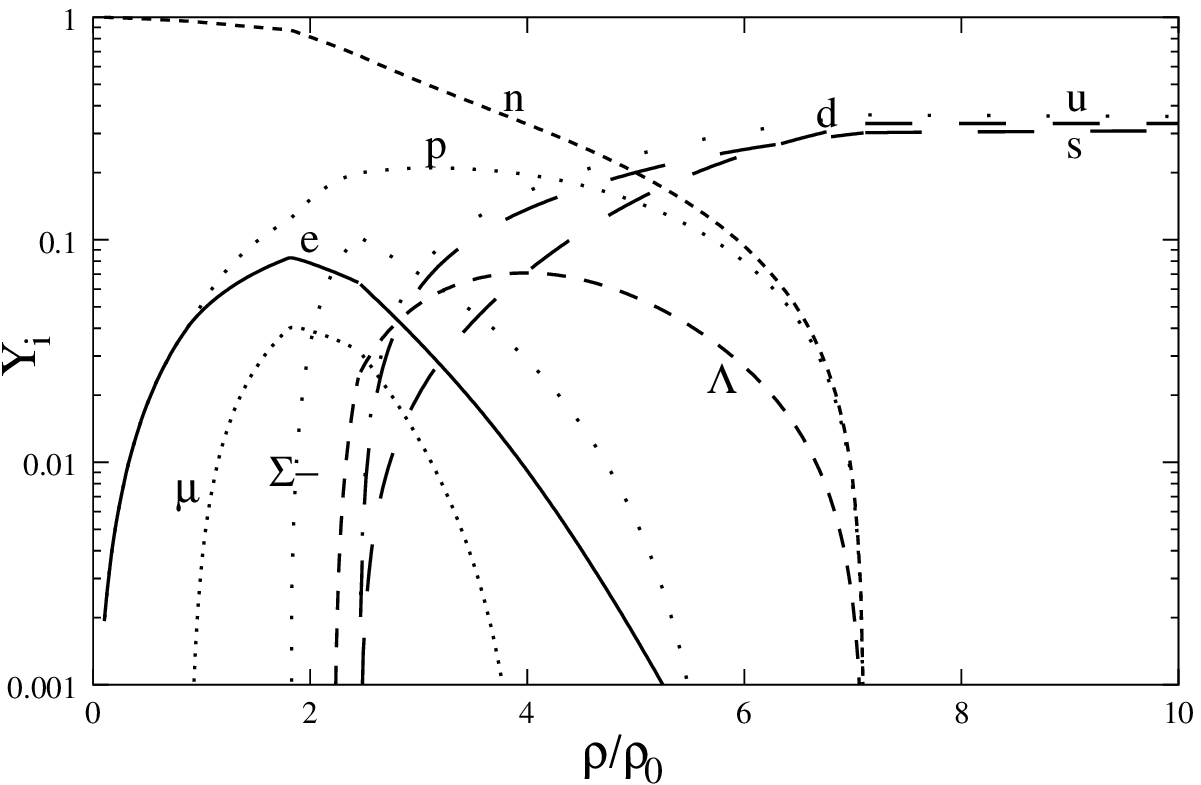}&
\includegraphics[width=7.0cm]{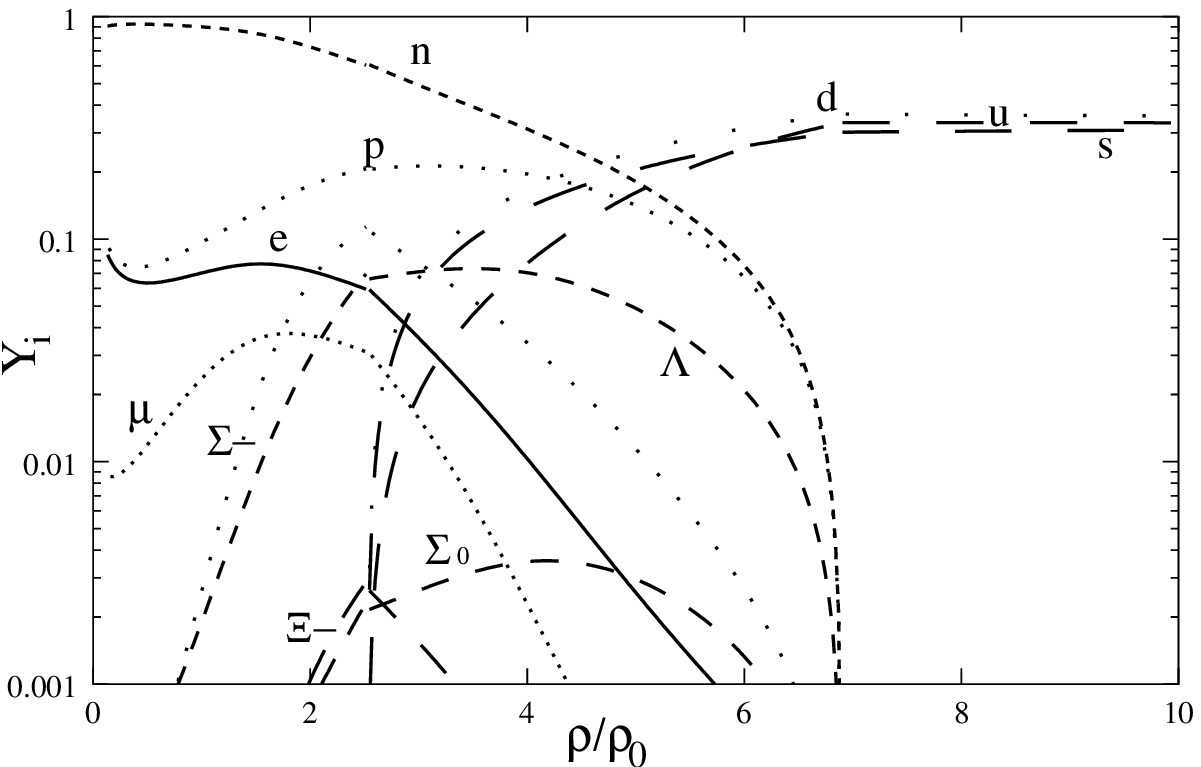}\\
a)&b)
\end{tabular}
\end{center}
\caption{Particle fractions $Y_i=\rho_i/\rho$, for $i=$ baryons, leptons and quarks, obtained with the GL force plus the Bag model 
for Bag$^{1/4}$=190 MeV and  a) T=0 b) T=20 MeV. In both cases 
$x_H=\sqrt{2/3}$.}
\label{fracoesbag}
\end{figure}

\begin{figure}
\begin{center}
\begin{tabular}{cc}
\includegraphics[width=7.0cm]{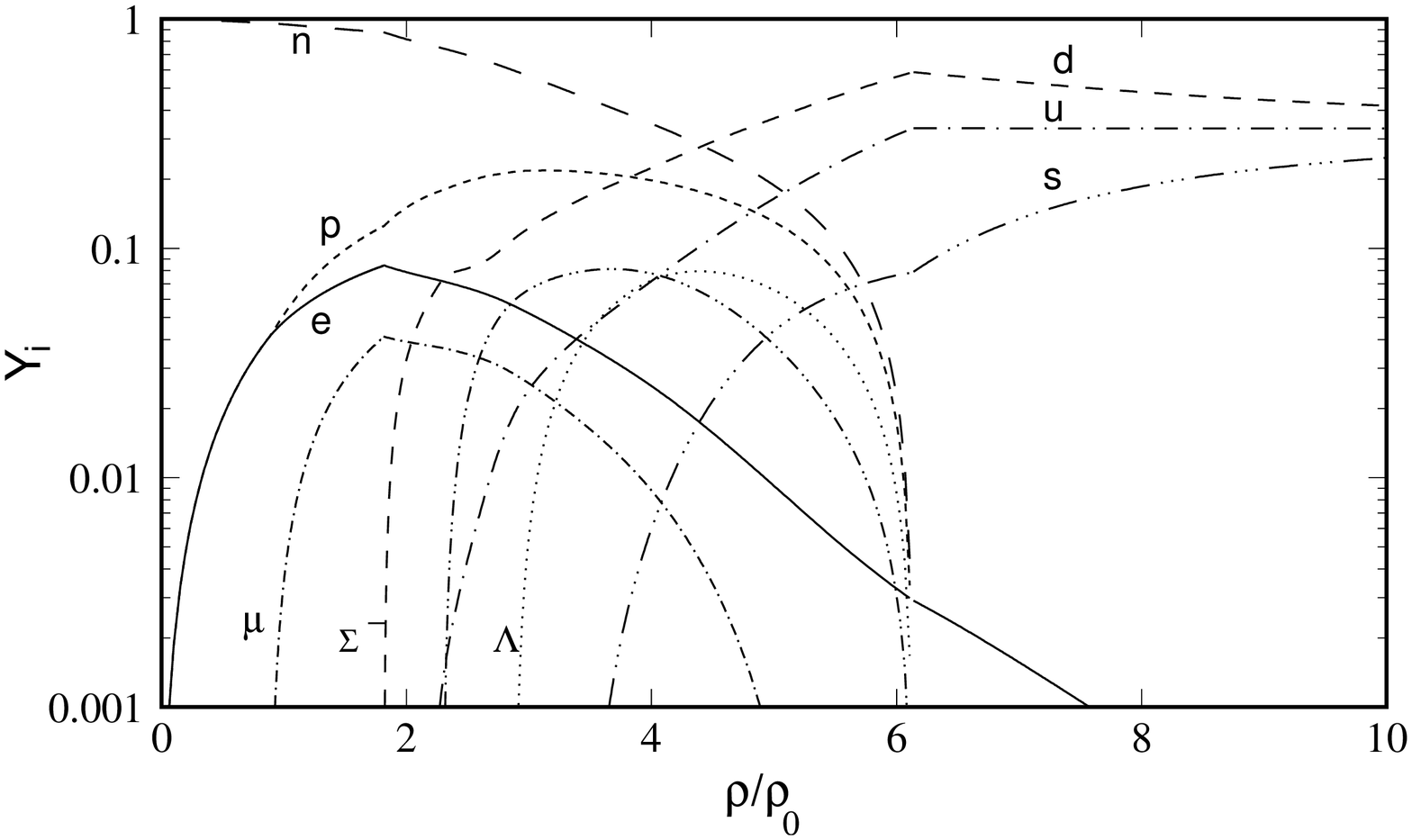}&
\includegraphics[width=7.0cm]{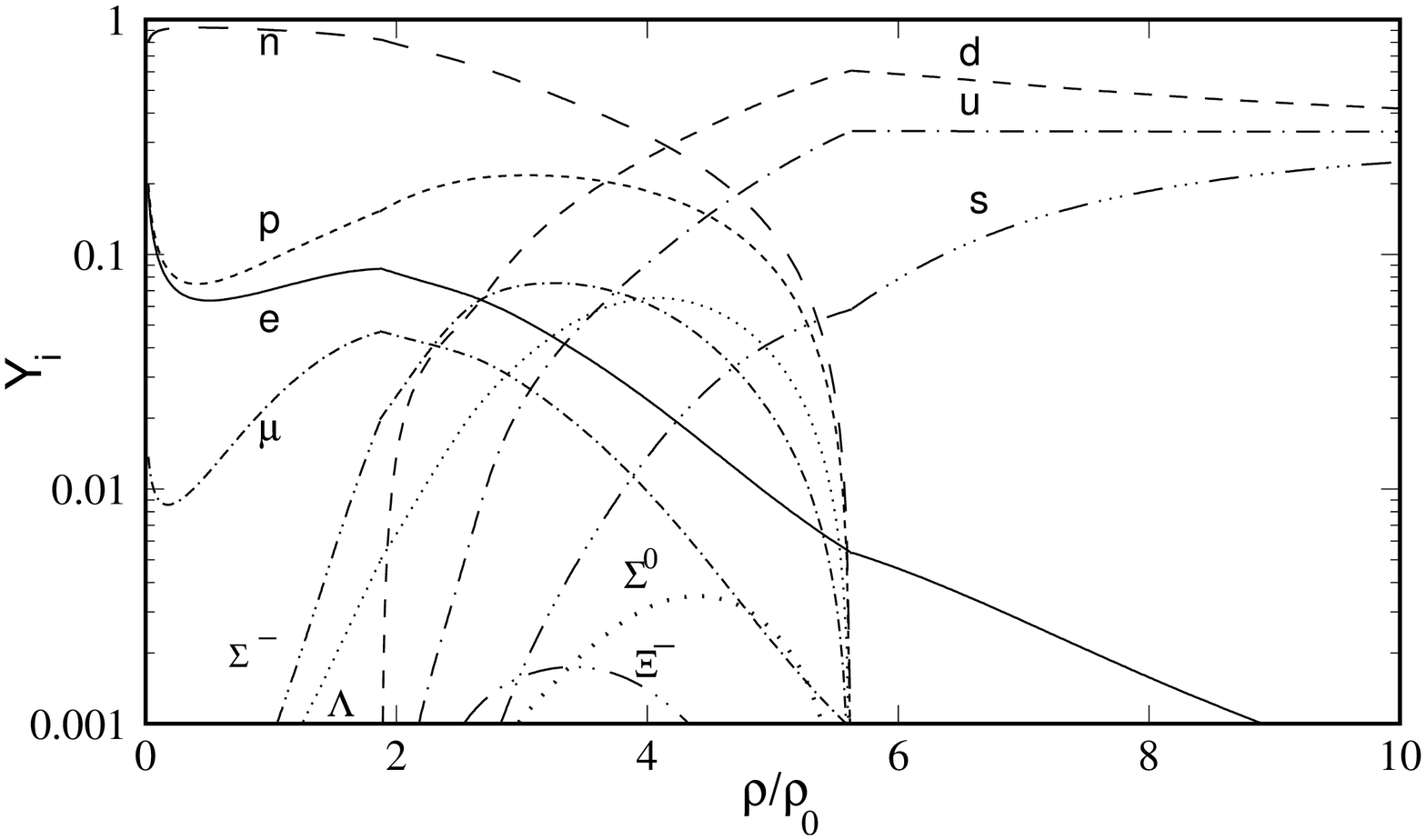}\\
a)&b)
\end{tabular}
\end{center}
\caption{The same as in figure \ref{fracoesbag} obtained with the NJL model and $x_\sigma=0.7$,
  $x_\omega=x_\rho=0.783$ for a) $T=0$ MeV; b) $T=20$ MeV.}
\label{fracoesnjl}
\end{figure}

\begin{figure}
\begin{center}
\includegraphics[width=7.cm]{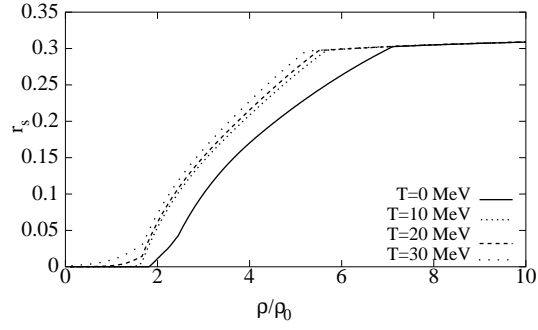}
\end{center}
\caption{Strangeness fraction $r_s$ for the EOS with the Bag model for the quark phase with $B^{1/4}=190$ MeV and $x_H=\sqrt{2/3}$.}
\label{cargabag}
\end{figure}

\begin{figure}
\begin{center}
\begin{tabular}{cc}
\includegraphics[width=7.0cm]{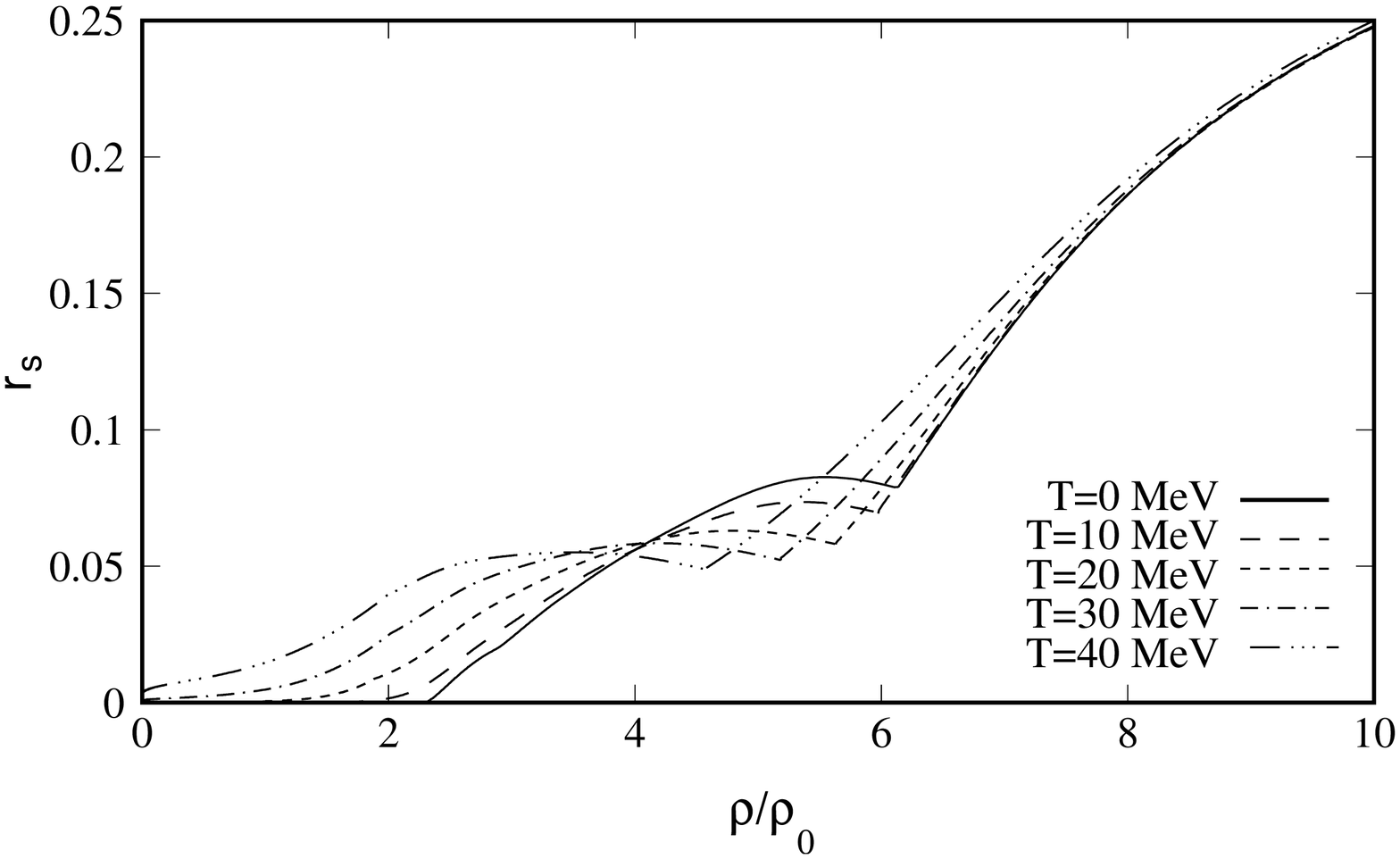}&
\includegraphics[width=7.0cm]{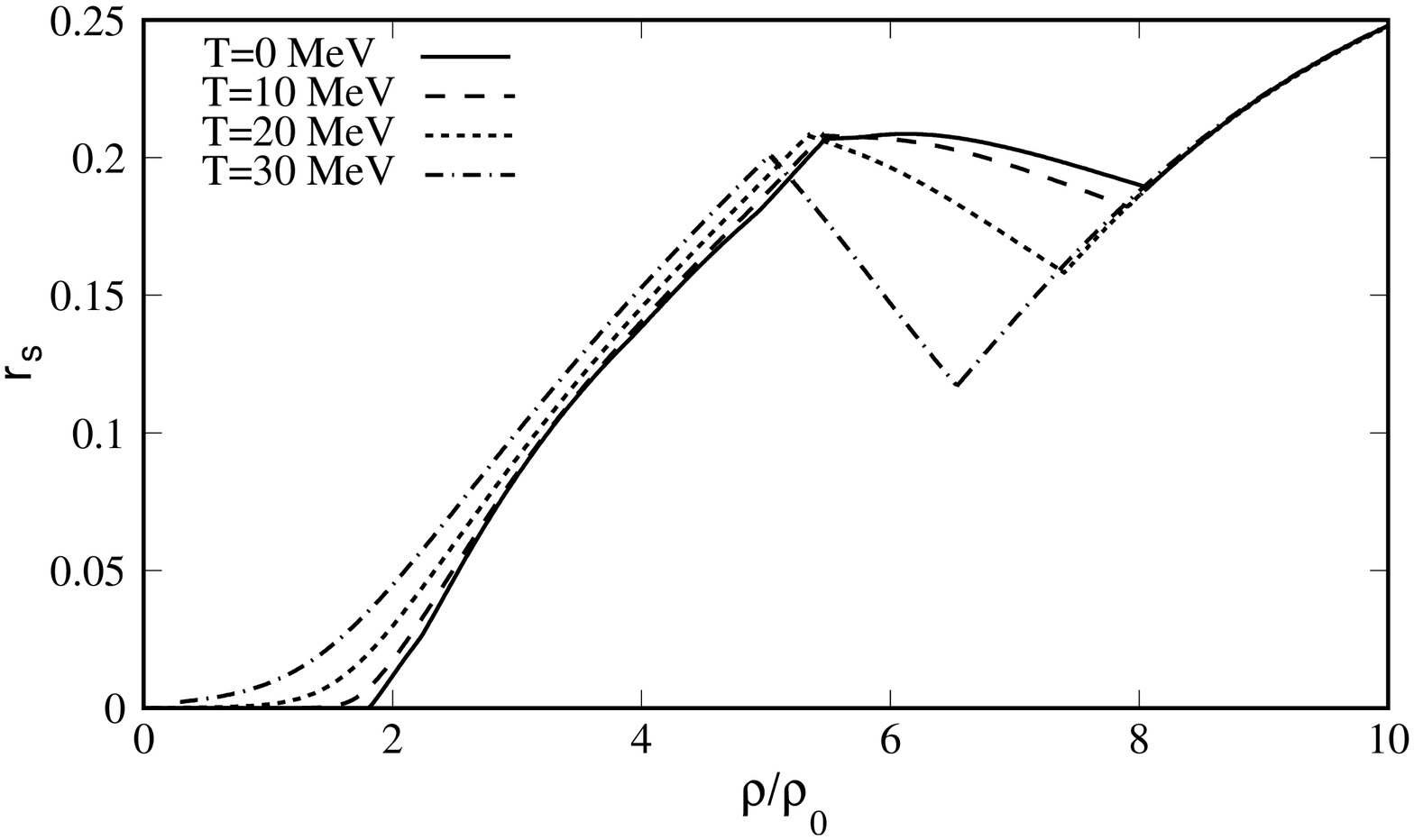}
\end{tabular}
\end{center}
\caption{Strangeness  fraction $r_s$ for the EOS with the NJL model for the quark phase: a)
  $x_\sigma=0.7$, $x_\omega=x_\rho=0.783$; b) $x_H=\sqrt{2/3}$.}
\label{carganjl}
\end{figure}

\end{document}